\documentclass[reprint,
amsmath,amssymb,
aps,
]{revtex4-2}
\usepackage{ulem,xcolor}
\usepackage{graphicx}
\usepackage{dcolumn}
\usepackage{bm}

\begin{document}
	
	\preprint{APS/123-QED}
	
	\title{Reservoir computing with logistic map}

	\author{R. Arun$^{1} $, M. Sathish Aravindh$^{1}$, A. Venkatesan$^{2}$, M. Lakshmanan$^{1}$} %
	\affiliation{$^{1}$Department of Nonlinear Dynamics, School of Physics, Bharathidasan University, Tiruchirappalli - 620 024, India. \\ $^{2}$ PG \& Research Department of Physics, Nehru Memorial College (Autonomous), Affiliated to Bharathidasan University, Puthanampatti, Tiruchirappalli - 621 007, India.}
	\email{arunbdu@gmail.com, \\ sathisharavindhm@gmail.com, \\ av.phys@gmail.com,\\ lakshman.cnld@gmail.com} 
	
	
	
	
	
\begin{abstract}
Recent studies on reservoir computing essentially involve a high dimensional dynamical system as the reservoir, which transforms and stores the input as a higher dimensional state, for temporal and nontemporal data processing. We demonstrate here a  method to predict temporal and nontemporal tasks by constructing virtual nodes as constituting a reservoir in reservoir computing using a nonlinear map, namely the logistic map, and a simple finite trigonometric series. We predict three nonlinear systems, namely Lorenz, R\"ossler, and Hindmarsh-Rose, for temporal tasks and a seventh order polynomial for nontemporal tasks with great accuracy. Also, the prediction is made in the presence of noise and found to closely agree with the target. Remarkably, the logistic map performs well and predicts close to the actual or target values. The low values of the root mean square error confirm the accuracy of this method in terms of efficiency. Our approach removes the necessity of continuous dynamical systems for constructing the reservoir in reservoir computing. Moreover, the accurate prediction for the three different nonlinear systems suggests that this method can be considered a general one and can be applied to predict many systems. Finally, we show that the method also accurately anticipates the time series of the all the three variable of R\"ossler system for the future (self prediction).
\end{abstract}
	
	\maketitle
	
	\section{Introduction}
Neural network (also known as artificial neural network (ANN)), deep neural network (DNN), and recurrent neural network (RNN) are subsets of machine learning, and they resemble human brain by mimicking the way how biological neurons process information within themselves \cite{cheng,schmid,rume}.
Reservoir computing (RC) which is a RNN based frame-work, is efficient in training to perform a given task\cite{luko,maass,jeager}. In RC, the network that is the hidden layer is not trained but only the output or readout layer is trained. 
The  computational effort is considerably reduced in RC when compared to RNN and this makes it a highly efficient and practical approach for various machine learning tasks \cite{gauth}.  Reservoir computing has been successfully applied in a wide range of applications, including time series prediction \cite{chen}, spatiotemporal prediction \cite{pathak}, speech recognition \cite{torre}, image classification \cite{tong}, etc. Also, chaotic signals that are overlaid are separated using RC.  Remarkably, the reservoir need not contain a large complex network and it can be any kind of dynamical system, as long as it is capable of transforming an input into higher dimensional state and as long as it can be perturbed by input and its output observed. The dynamical system which is rich in its dynamics can act like a high-dimensional system and encode the inputs as a temporal pattern instead of multiplexing the input in state space as in conventional RC. In 2011, Appeltant $et~al$ showed that inputs can be multiplexed in time rather than space by turning a single nonlinear node as a virtually high dimensional system \cite{appel}. By constructing a reservoir in terms of virtual nodes they have demonstrated speech recognition task.   Dion, Mejaouri and Sylvestre have achieved RC by constructing a network of virtual nodes multiplexed in time by the dynamics of oscillating silicon beam which exhibits Duffing nonlinearity and proved time series prediction and spoken word recognition \cite{dion}. Haynes $et~al$ have shown that a single autonomous Boolean logic element can be used as a physical system \cite{haynes}. Jensen and Tufte have employed Murali-Lakshmanan-Chua chaotic circuit to provide the nonlinear node and predicted nontemporal tasks \cite{jensen}. Very recently, Mandal, Sinha and Shrimali have proved that a single forced driven pendulum can be considered as a single node reservoir in RC and they have predicted the temporal and nontemporal tasks with great accuracy \cite{sinha}.
	
In this article, following the general methodology for RC \cite{appel, dion, haynes, jensen, sinha} where the reservoir can be constructed by the virtual nodes, we propose the interesting possibility that one can use as RC just a nonlinear map or the map with a simple trigonometric function, encoding the inputs as temporal patterns instead of a continuous dynamical system. Also, we prove the universality of this method by predicting both nontemporal and temporal tasks. In the case of a nontemporal task, we predict the nature of a polynomial function, and in a temporal task, we predict three nonlinear chaotic systems that include Lorenz, R\"ossler and Hindmarsh-Rose systems with great accuracy. 
Also, we train the reservoir and predict the time-series of the variables $x$, $y$ and $z$ of the R\"ossler system by supplying $x$, $y$ and $z$, respectively, as inputs (closed loop prediction).
	
The paper is organized as follows: In Sec. II, the prediction for the nontemporal task is discussed to predict a polynomial function, and in Sec. III, the temporal task is explained to predict three nonlinear systems, namely the Lorenz, R\"ossler and Hindmarsh-Rose oscillators using the logistic map as a reservoir for the prediction.  We have also predicted the closed-loop task for R\"ossler system and investigated the performance of the RC against hyper-parameters. In Appendix A, both the nontemporal and temporal tasks are predicted out by incorporating an additional simple trigonometric series function as a reservoir, and in Appendix B the dynamical behaviour of the logistic map is briefly discussed for ready reference. Finally, we present our conclusion in Sec. V.
	
\section{Prediction for nontemporal Task}
For a nontemporal task we consider the prediction of a seventh degree polynomial $f(x)$ = $(x-3)(x-2)(x-1)x(x+1)(x+2)(x+3)$ for $u_{min}\leq x \leq u_{max}$.  From a few available samples $(x_i, f(x_i))$ ($=(u_i,v_i)$ for convenience) between the chosen limits $u_{min}$ and $u_{max}$, where $i$ = 1,2,3,...,$L$, a state vector matrix $R$ of the reservoir is formed and then a weight matrix $W$ corresponding to the linear transformation between the actual output and the reservoir state vector matrix is determined by training the reservoir. Further, the polynomial is predicted for the complete set of $x$ values between $u_{min}$ and $u_{max}$ by using the obtained weight matrix.
\begin{figure}
\centering
\includegraphics[width=1\linewidth]{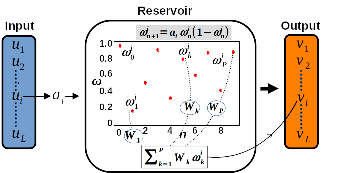}
\caption{The schematic diagram for training with the logistic map. The input $u_i$ is linearly transformed into $a_i$ and then is supplied into the logistic map to form virtual nodes for the reservoir. $V_i$, $i=1,2,..,L$ is the output generated from the reservoir. The iterated values of the logistic map ($\omega_k^i,~k=1,2,...,P$) are indicated as red colour dots. The $W_k,~k=1,2,...,P$, are the components of the weight matrix $W$.  The black colour dotted lines inside the reservoir indicate the internal computations of the reservoir.}
\label{res}
\end{figure}
Two row vectors $u$ and $v$ of size 1$\times L$ are formed from the samples $(u_i,v_i)$ as ${u}$ = $[u_1,u_2,...,u_i,...,u_{L}]$ and ${v}$ = $[v_1,v_2,...,v_i,...,v_{L}]$. The state vector matrix of the reservoir is constructed by $L$ column vectors corresponding to each input $u_i$, $i$ = 1,2,...,$L$. Each input is multiplexed, forming virtual nodes, and state vector of size $L\times 1$ corresponding to each input is formed.  For multiplexing, each input element $u_i$ is supplied to the logistic map equation 
	\begin{align}
		\omega_{n+1} = a_i ~\omega_n~ (1 - \omega_n),~~~i=1,2,..L, \label{logistic}
	\end{align}
	via $a_i$ through the linear transformation
	\begin{equation}
		a_i = a_{min}+ \left(\frac{a_{max} - a_{min}}{u_{max}-u_{min}}\right)(u_i - u_{min}) \label{a}.
	\end{equation}
	The quantities $a_{min}$ and $a_{max}$ in Eq.\eqref{a} can be chosen accordingly for the best prediction and also they can be chosen based on the requirement of using chaotic or non-chaotic region, in the bifurcation diagram of the logistic map (\ref{logistic}), for better prediction (see Fig.\ref{bifur}). Equation \eqref{logistic} is iterated for $\omega_1$, $\omega_2$,..., $\omega_P$ corresponding to each $a_i$. The initial value of $\omega$ = $\omega_0$ is taken as 0.95 throughout this paper. $P$ is a hyper-parameter i.e. it can be tuned to enhance the accuracy in prediction.  Therefore, we can generate $P$ number of $\omega$'s for each $a_i$ as $\omega_1^i,\omega_2^i,...,\omega_P^i$.  The input $u_i$ is then multiplexed via $a_i$ by $\omega_1^i,\omega_2^i,...,\omega_P^i$. A reservoir state vector(column vector) ${Y}_{i}$ corresponding to $a_i$ with size ($P\times 1$) is formed from the data points $\omega_1^i,\omega_2^i,...,\omega_P^i$ as 
	\begin{align}
		{Y}_{i} = [ \omega^{i}_1,\omega^{i}_2,....,\omega^{i}_P]^T. \label{Y}
	\end{align}
	Similarly, the reservoir state vectors $Y_{1},Y_{2},....,Y_{L}$ for $L$ inputs are computed corresponding to the choice of parameters $a_1,~a_2,~a_3,....,a_L$, respectively, and they are stacked together to from a reservoir state vector matrix $R = [Y_{1},Y_{2},....,Y_{L}]$ with the size $P\times L$ as
	\begin{align}
		R = 
		\begin{bmatrix}
			\omega_{1}^1 & \omega_{1}^2 & \omega_{1}^3 & ......&\omega_{1}^L\\~\\
			\omega_{2}^1 & \omega_{2}^2 & \omega_{2}^3 & ......&\omega_{2}^L\\~\\
			\omega_{3}^1 & \omega_{3}^2 & \omega_{3}^3 & ......&\omega_{3}^L\\
			...&...&...&...&...\\
			...&...&...&...&...\\
			\omega_{P}^1 & \omega_{P}^2 & \omega_{P}^3 & ......&\omega_{P}^L
		\end{bmatrix}. \label{Rnt}
	\end{align}
	
	The linear transformation between the vector $v$ and the reservoir state vector matrix $R$ is given by (see Fig.\ref{res})
	\begin{align}
		v = W_{n} R, \label{vWR}
	\end{align}
	where $W_{n}$ is the weight matrix for the nontemporal $ (n) $ task case with size (1$\times$P) and it can be obtained from the reservoir state vector matrix $R$ and the actual output vector $v$ by Eq.\eqref{vWR} as
	\begin{align}
		W_{n} = vR^{-1}, \label{W}
	\end{align}
	by employing the Moore-Penrose pseudoinverse \cite{barata}.  
	
Using the above weight matrix $W_n$, the polynomial $f(x)$ can be predicted for $L$ new inputs $u' = [u_1',u_2',...,u_L']$ for which the output is unknown by
\begin{align}
V = W_{n} R'. \label{VWR}
\end{align}
The $L$ elements of $u'$ can be random or uniform between $x=u_{min}$ and $x=u_{max}$. The matrix $R'$ in Eq.\eqref{VWR} is the reservoir state vector matrix constructed for $u'$ by the same procedure discussed above. We predict the polynomial $f(x)$ for $L$(=100) number of $x$'s uniformly distributed between $u_{min}$ = -3 and $u_{max}$ = +3 and it is plotted in Fig.\ref{fig1}(a) in the absence of noise. The parameters are kept as $P$ = 100, $a_{min}$ = 2.1 and $a_{max}$ = 2.2.  The solid blue line indicates the actual result and the red open circles indicate the prediction that matches well with the actual result.
\begin{figure}
\centering
\includegraphics[width=1\linewidth]{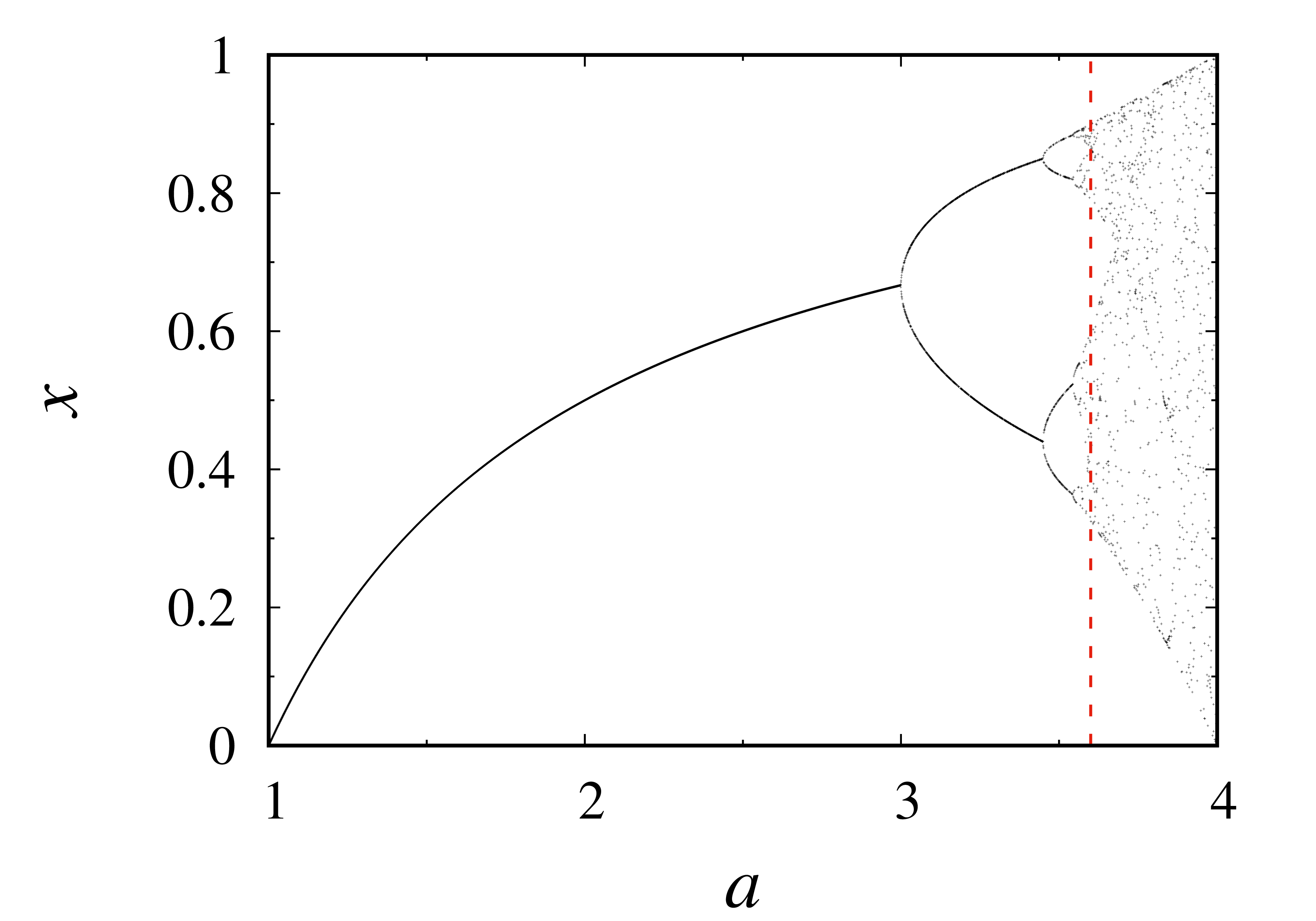}
\caption{Bifurcation diagram of the logistic map. The vertical dashed line separates the periodic region and chaotic region ($a > 3.57$). The periodic regions include period-1 region ($1.0 < a < 3.0$), period-2 cycle region ($3.0 < a < 3.449$), period-4 cycle region ($3.449< a< 3.544112$) and so on (see for example, ref.\cite{lak}).}
\label{bifur}
\end{figure}
To verify the efficiency of this procedure in prediction over noise, both the input and output of the the samples ($u_i,v_i$) are supplied along with the strength of noise $\delta$ = 0.1 (i.e. i.e. the random noise is generated between -$\delta$ and +$\delta$ using uniform distribution function.).  The weight matrix $W_n$ is then obtained corresponding to these noisy samples and then the prediction is made for the uniformly distributed $x$'s (along with noise) between -3 and +3 as discussed before. In Fig.\ref{fig1}(b), the predicted output plotted by red downward triangles match well with the target values plotted by the blue upward triangles. In order to confirm the prediction in the chaotic region corresponding to $a_{min}$ = 3.8 and $a_{max}$ = 3.9, the polynomial is predicted for $P$ = 20 with the strength of the noise being 0.1 and the result is plotted  in Fig.\ref{fig1}(c).
\begin{figure}
\centering
\includegraphics[width=0.9\linewidth]{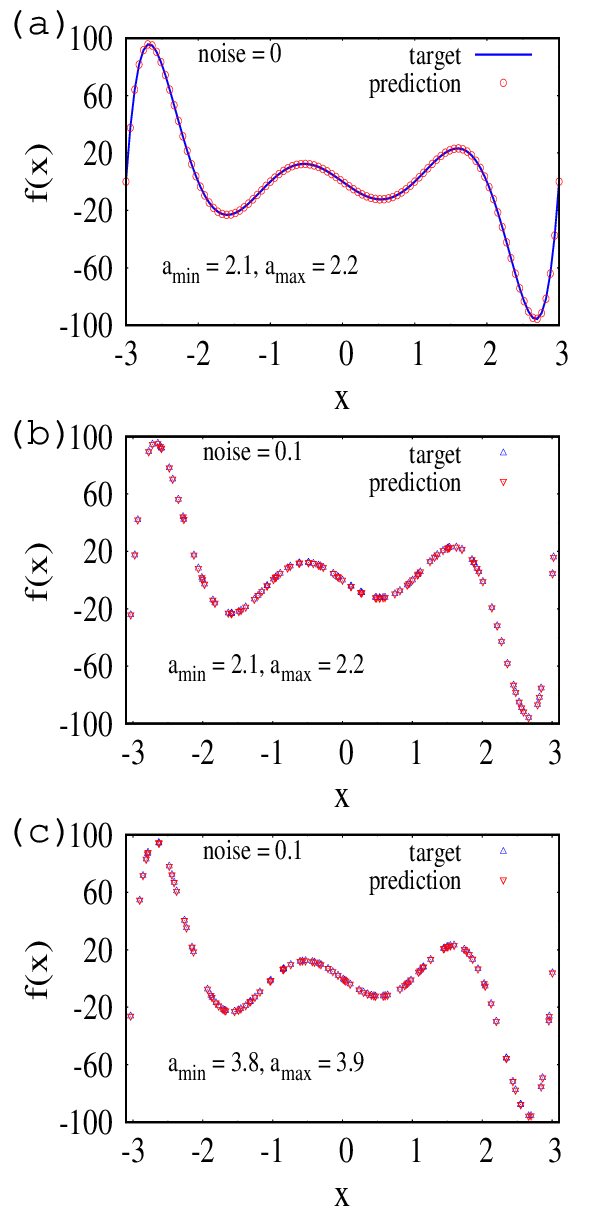}
\caption{ Prediction of the polynomial $f(x)$ = $(x-3)(x-2)(x-1)x(x+1)(x+2)(x+3)$ for $ x $ between -3 and +3 for (a) $a_{min}$ = 2.1, $a_{max}$ = 2.2, $P$ = 100, $\delta$ = 0, (b) $a_{min}$ = 2.1, $a_{max}$ = 2.2, $P$ = 100, $\delta$ = 0.1 and (c) $a_{min}$ = 3.8, $a_{max}$ = 3.9, $P$ = 20, $\delta$ = 0.1. Here $L$ = 100, $u_{min}$ = -3 and $u_{max}$ = 3.}
\label{fig1}
\end{figure}
From Figs.\ref{fig1} we can clearly observe that the polynomial is accurately predicted in both the cases of absence and presence of noise, for the bifurcation parameter $a$ in the chaotic as well as the nonchaotic regions. Statistically, the accuracy of the prediction with the target is measured by the root mean square error
\begin{align}
RMSE = \sqrt{\sum_{i=1}^{L}{(v_i - V_i)^2}/{L}}, \label{rmse}
\end{align}
and normalized root mean square error
\begin{align}
NRMSE = \frac{RMSE}{\sum_{i=1}^{L}{(V_i - V_{avg})^2}/{L}}. \label{nrmse}
\end{align}

The set $v_i$ is the actual output value set for the corresponding predicted output set $V_i$. The $V_{avg}$ is the mean of the output set $V_i$. In the above nontemporal task for predicting the polynomial $f(x)$ the values of (RMSE, NRMSE) are calculated as (3.475412$\times$10$^{-3}$, 2.305818$\times$10$^{-6}$) and (6.152246$\times$10$^{-2}$, 3.976396$\times$10$^{-5}$) corresponding to the absence (Fig.\ref{fig1}(a)) and presence (Fig.\ref{fig1}(b)) of noise, respectively, for the nonchaotic region and (6.689998$\times$10$^{-2}$, 4.1269$\times$10$^{-5}$) corresponding to the chaotic region (Fig.\ref{fig1}(c).

Here, one may note that the prediction for $f(x)$ has been done with $a_{min}$ = 2.1 and $a_{max}$ = 2.2 in the nonchaotic region and $a_{min}$ = 3.8 and $a_{max}$ = 3.9 in the chaotic region. The input is supplied through the small window (i.e. $\triangle a=a_{max}-a_{min}$ = 0.1) of the parameter $a$ of the logistic map. The parameter window is small and cannot be increased more than 0.1 while maintaining the RMSE below 1.0 for better prediction.  
\begin{figure}
\centering
\includegraphics[width=0.9\linewidth]{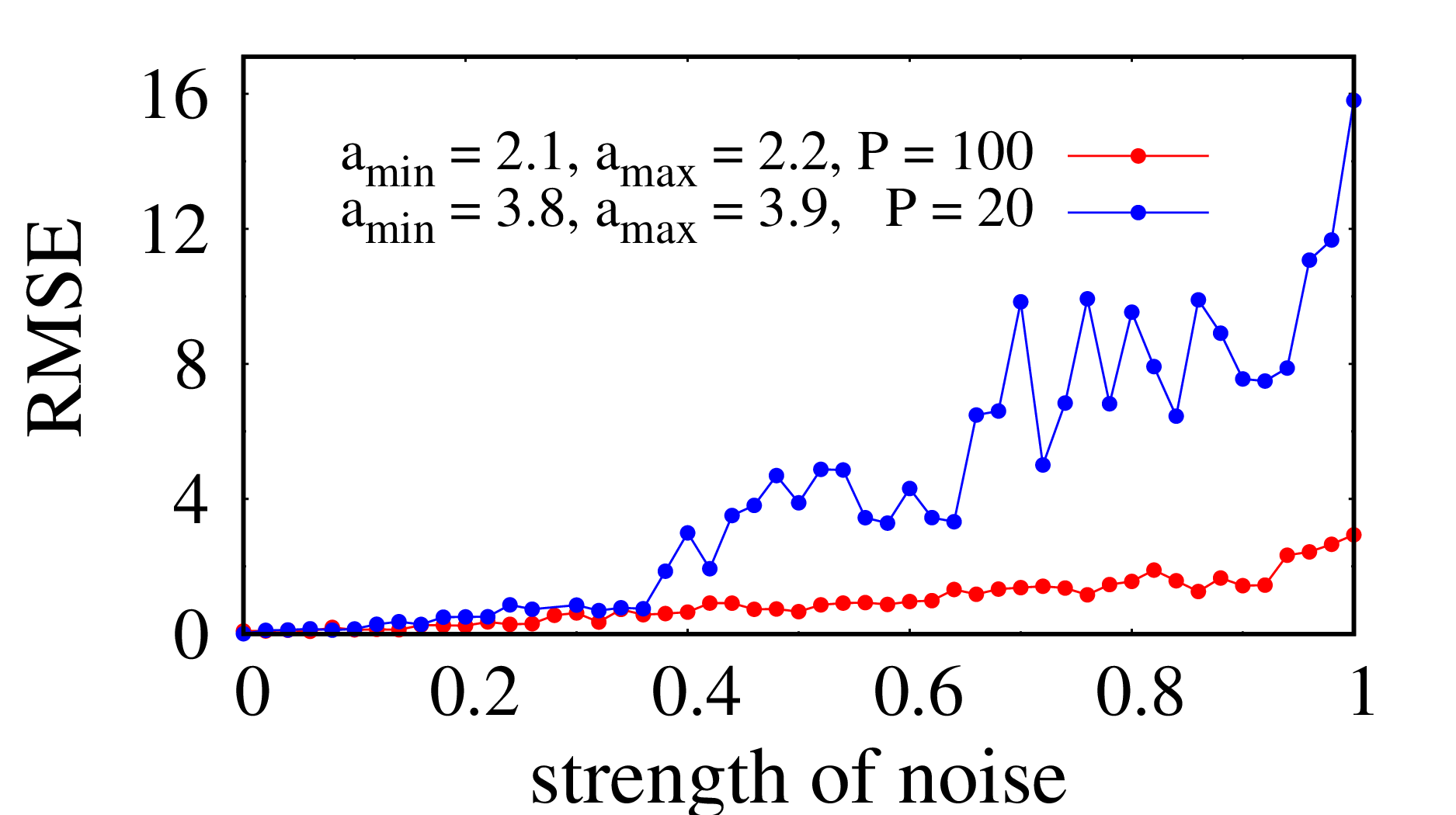}
\caption{Variation of RMSE with respect to the strength of noise for the nonchaotic (red) and chaotic (blue). Here, $u_{min}$ = -3.2 and $u_{max}$ = 3.2.}
\label{noiseRMSE}
\end{figure}

We also note that the parameter window can be increased and the polynomial function can be predicted over a wider range of $a$, including the chaotic regime, by forming a simple finite series of trigonometric functions from the values obtained through the logistic map ($\omega_i,~i=1,2,...,P$). The details are given in Appendix A.

Further, the performance of the reservoir over the strength of noise in predicting the polynomial is investigated by plotting the RMSE against the strength of noise for $a_{min}$ = 2.1 and $a_{max}$ = 2.2 while $P$ = 100 and for $a_{min}$ = 3.8 and $a_{max}$ = 3.9 while $P$ = 20 in Fig.\ref{noiseRMSE}. It clearly shows that when the logistic map operates in chaotic and nonchaotic region, the value of RMSE is maintained below 1.0 as long as the strength of the noise is below 0.36 and 0.62, respectively, and beyond these strengths the error increases considerably.
\section{Prediction for Temporal task}
In this section we intend to perform the temporal task of predicting the time series of one variable of a nonlinear system by supplying the time series of another variable of the same system as input.  We consider three different nonlinear systems, namely the Lorenz, R\"ossler and Hindmarsh-Rose oscillators, and predict their $y(t)$ variable from that of $x(t)$. This temporal task of predicting the time series requires a sequential form of input of the past data to predict the future. For this temporal task, let the total number of samples available for training is $L$ and the samples are given by ($x_i(t),y_i(t)$), where $i$ = 1,2,...,$L$.  For convenience let ($u_i,v_i$) = ($x_i(t),y_i(t)$). Unlike the nontemporal task that we discussed in the previous section, these samples $(u_i,v_i)$ are sequential data corresponding to time $t$ = $t_1,t_2,t_3$,...,$t_L$.  Here the $u_i$ and $v_i$ can be denoted as inputs and outputs, respectively, and their corresponding state vectors are given by the row vectors  $u=[u_1,u_2,.....,u_L]$ and $v=[v_1,v_2,.....,v_L]$.  The reservoir state vector ($Z_{i}$) corresponding to the input $u_i$ is constructed as $Z_{i} = [g_0 Y_{{i-m}},.........,g_{m-1} Y_{{i-1}},g_m Y_{i}]^T$ along with $m$ previous inputs to bring the effect of past memory in prediction. Here, the finite memory parameter $m$ is a hyper parameter. Also $Y_k$, $k = i-m,i-m+1,.....,i$, corresponding to the input $u_k$, $k = i-m,i-m+1,.....,i$, is similarly constructed as a column vector with size $(P \times 1)$ by Eq.\eqref{Y}, using Eqs.\eqref{logistic} and \eqref{a}, in the same way explained above for the nontemporal task. Here $g_j$, $j$ = 1,2,..,$m$ are the weights assigned to $Y_k$, $k = i-m,i-m+1,.....,i$.  The values of $g_j$ are uniformly distributed in the range  [0,1] and hence $g_0$ = 0 and $g_m$ = 1. Therefore, the size of the reservoir state vector $Z_{i}$ for the temporal task  is $(mP\times 1)$. Similarly, all the reservoir state vectors $Z_{1}$, $Z_{2}$,...,$Z_{L}$ corresponding to the inputs $u_i$, $i=1,2,...,L$, respectively, are obtained and the reservoir state vector matrix $R$ is formed by staking them together as $R = [Z_1,Z_2,.....,Z_L]$. Since $g_0 =  0$ the reservoir state vector matrix $R$ is constructed for the samples $(u_i,v_i)$, $i=1,2,...,L$ as
\begin{align}
R = 
\begin{bmatrix}
g_1 Y_{2-m}~ & g_1 Y_{3-m}~ & g_1 Y_{4-m} ~& ......&~g_1 Y_{L-m+1}\\~\\
g_2 Y_{3-m}~ & g_2 Y_{4-m}~ & g_2 Y_{5-m}~ & ......&~g_2 Y_{L-m+2}\\~\\
g_3 Y_{4-m}~ & g_3 Y_{5-m}~ & g_3 Y_{6-m}~ & ......&~g_3 Y_{L-m+3}\\
...&...&...&...&...\\
...&...&...&...&...\\
...&...&...&...&...\\
g_{m-2} Y_{-1}~ & g_{m-2} Y_0~ & g_{m-2} Y_1~ & ......& ~g_{m-2} Y_{L-2}\\
g_{m-1} Y_0~ & g_{m-1} Y_1~ & g_{m-1} Y_2 ~& ......& ~g_{m-1} Y_{L-1}\\
g_m Y_1~ & g_m Y_2~ & g_m Y_3~ & ......&~ g_m Y_L
\end{bmatrix}\label{Rt}
\end{align}
with the size of $(mP \times L)$. Using the above matrix $R$ the weight matrix $W_t$ for this temporal $ (t) $ task can be obtained as
\begin{align}
W_t = v R^{-1} \label{tempW}.
\end{align}
The size of the weight matrix $W_t$ is $(1\times mP)$. After obtaining the weight matrix $W_t$, the outputs corresponding to the new set of sequential data inputs $u' = [u_2,u_3,.....,u_{L+1}]$ are predicted for the output $V = [V_2, V_3,.....,V_{L+1}]$ using the relation
\begin{align}
V = W_t R' \label{tempV}.
\end{align}
Here $R'$ is the revised reservoir state vector matrix for the new set of inputs. Similarly, the process is repeated to predict the outputs for the successive set of sequential inputs. 
	
\subsection{Lorenz system}
\begin{figure}
\centering
\includegraphics[width=1\linewidth]{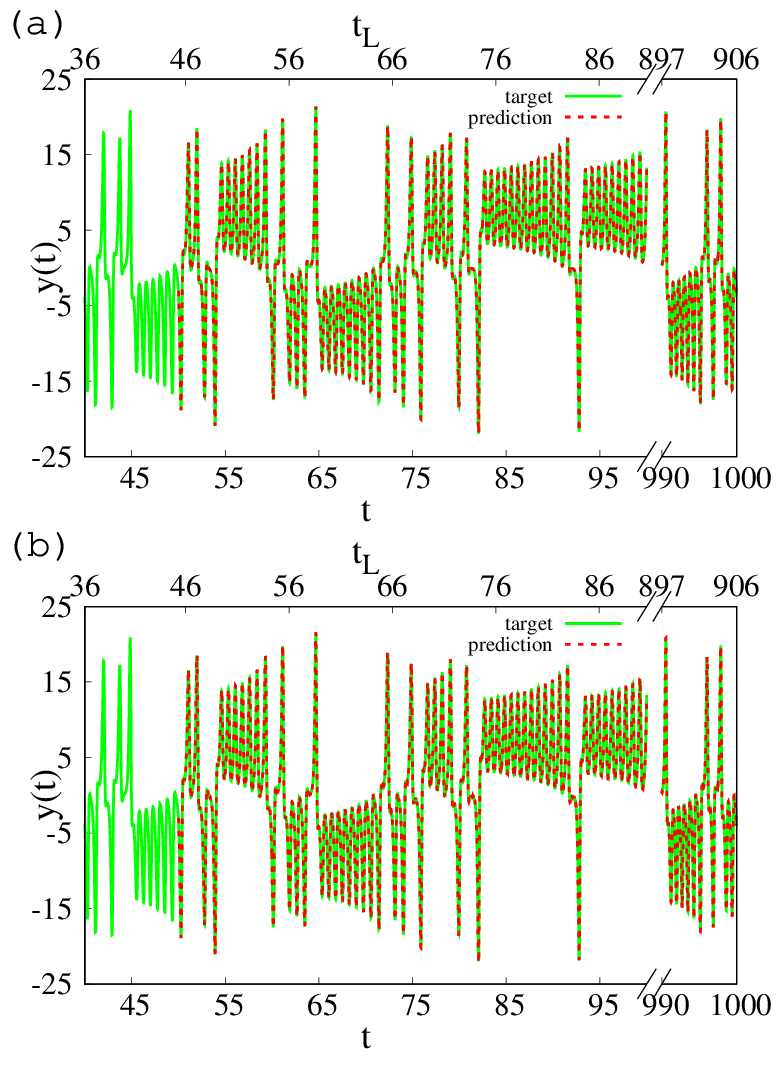}
\caption{Prediction of Lorenz system for $y(t)$ from the input $x(t)$ for the strengths of (a) $\delta$ = 0 and (b) $\delta$ = 0.01. Here, $a_{min}$ = 1, $a_{max}$= 2, $u_{min}$ = -17, $u_{max}$ = 17, $P$ = 3 and $m$ = 100. In each of the figures, we have also indicated the Lyapunov time $t_L = t\cdot\lambda_{max}$, where $\lambda_{max}$ = 0.906 is the maximal Lyapunov exponent of the Lorenz system \cite{sprott}.}
\label{lorenz}
\end{figure}
\begin{figure}
\centering
\includegraphics[width=1\linewidth]{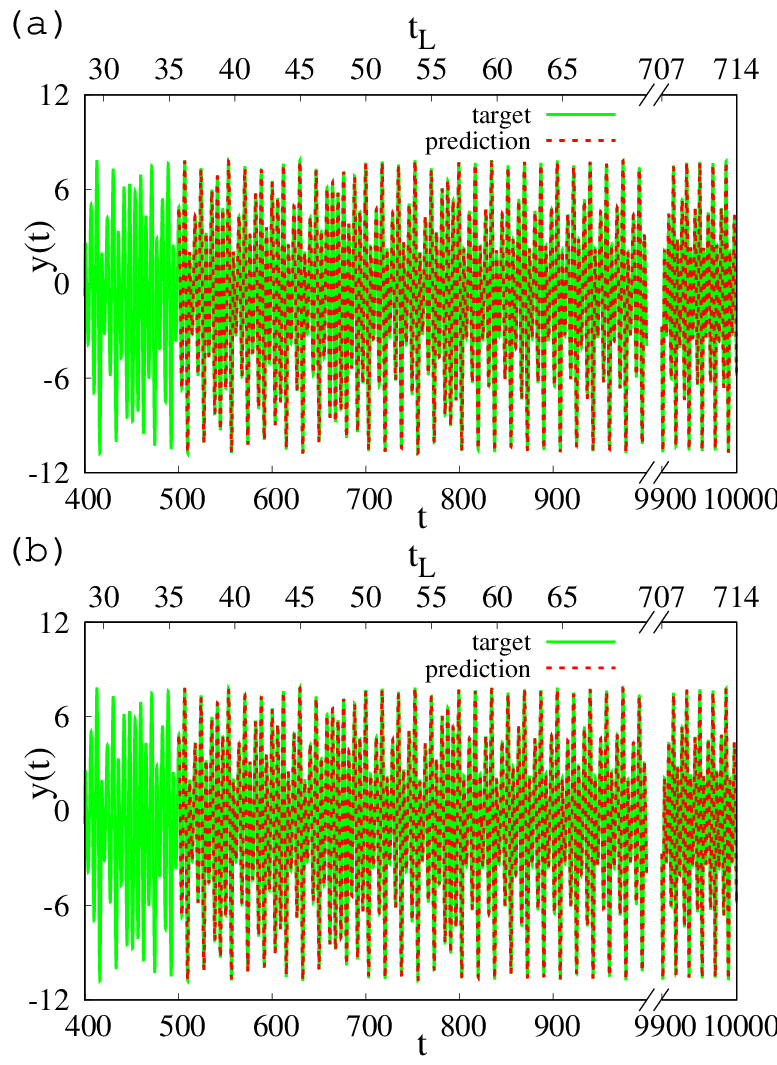}
\caption{Prediction of R\"ossler system for $y(t)$ from the input $x(t)$ for the strength of (a) $\delta$ = 0 and (b) $\delta$ = 0.01. Here, $a_{min}$ = 1, $a_{max}$= 2, $u_{min}$ = -10, $u_{max}$ = 12, $P$ = 3, $m$ = 100 and $\lambda_{max}$ = 0.0714 \cite{sprott}.}
\label{rossler}
\end{figure}
The Lorenz system, governed by the following equations 
\begin{eqnarray}
\dot{x} &=& 10(y-x),\nonumber\\
\dot{y} &=& x(28-z)-y,\nonumber\\
\dot{z} &=& x y - (8/3)z,
\end{eqnarray}
is predicted for the variable $y(t)$ from the input $x(t)$. The predictions are shown from $t$ = 50 to $t$ = 100 in Figs.\ref{lorenz}(a) and \ref{lorenz}(b) in the absence and presence of noise, respectively, where the actual output i.e. target is plotted in green solid line and the prediction is plotted as red dashed line. The strength of the noise has been taken as 0.01. The prediction is made with the choice of the parameters as $a_{min}$ = 1, $a_{max}$= 2, $u_{min}$ = -17, $u_{max}$ = 17, $P$ = 3 and $m$ = 100. The training is done from $t$ = 40.00 to $t$ = 49.99 with 1000 data points and the prediction is made from $t$ = 50.00 to $t$ = 99.99 as shown in Figs.\ref{lorenz}(a) and (b). From Figs.\ref{lorenz}(a) and (b) we can observe that the prediction is good and matches well with the target.  The root mean square error RMSE is calculated for the prediction between the times $t$ = 50 and $t$ = 1000 (95000 numbers of data) as 1.541608$\times$10$^{-3}$ in the absence of noise and 7.168561$\times$10$^{-2}$ in the presence of noise.  The initial conditions for training as well as prediction are $(x^*,y^*,z^*)$ = (25, 18, 120) and the time step is kept as 0.01 for the generation of input $x(t)$ and prediction of output $y(t)$.  It has been verified that the prediction is made even after $t$ = 1000 with good accuracy. We have also indicated the (rescaled) Lyapunov time $t_L = t\cdot\lambda_{max}$, where $\lambda_{max}$ is the maximal Lyapunov exponent, in each of the figures \ref{lorenz} and also in Figures \ref{rossler} and \ref{hindmarsh} below  for other systems.
\subsection{R\"ossler system}
The R\"ossler system, governed by the equations
\begin{eqnarray}
\dot{x} &=& -y - z,\nonumber\\
\dot{y} &=& x + 0.2 y,\nonumber\\
\dot{z} &=& 0.2 + z(x - 5.7),
\end{eqnarray}
is predicted and plotted in Figs.\ref{rossler}(a) and \ref{rossler}(b) in the absence and presence of the noise, respectively for the parameters $a_{min}$ = 1, $a_{max}$= 2, $u_{min}$ = -10, $u_{max}$ = 12, $P$ = 3 and $m$ = 100. Here the strength of the noise is 0.01. In both the figures the target is plotted by solid green line and the prediction is plotted by dashed red line between the times $t$ = 500.0 and $t$ = 999.9. The training was made with 1000 data points between the times $t$ = 400.0 and $t$ = 499.9. Figs.\ref{rossler}(a) and (b) imply that the prediction is great for both the cases of absence and presence of noise.  The initial conditions $(x^*,y^*,z^*)$ = (0.1,0.2,0.3) are taken for the training and prediction. The time step is maintained as 0.1. The root mean square error for the predictions between $t$ = 500 and $t$ = 10000 with 95000 data in the absence and presence  of the noise are calculated as 4.660823$\times$10$^{-2}$ and 5.376042$\times$10$^{-2}$, respectively.
\subsection{Hindmarsh-Rose system}
\begin{figure}
\centering
\includegraphics[width=1\linewidth]{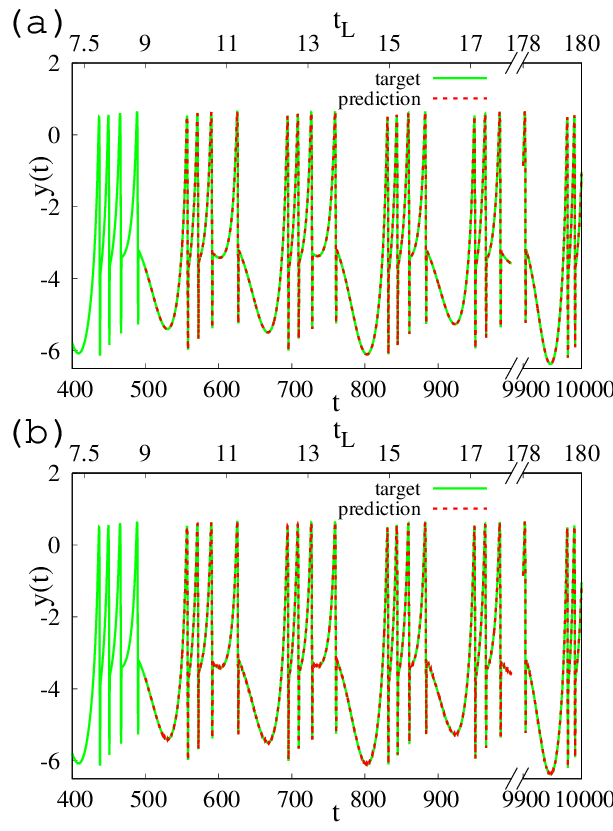}
\caption{Prediction of Hindmarsh-Rose system for $y(t)$ from the input $x(t)$ for the strength of (a) $\delta$ = 0 and (b) $\delta$ = 0.01. Here, $a_{min}$ = 1, $a_{max}$= 2, $u_{min}$ = -1.2, $u_{max}$ = 1.8, $P$ = 5, $m$ = 100 and $\lambda_{max}$ = 0.018 \cite{Huag}}.
\label{hindmarsh}
\end{figure}
The Hindmarsh-Rose (HR) system, governed by the equations 
\begin{eqnarray}
\dot{x} &=& y + 3x^2 -x^3 -z + 3.25,\nonumber\\
\dot{y} &=& 1 - 5x^2 - y,\nonumber\\
\dot{z} &=& 0.006(4(x+1.6) - z),
\end{eqnarray}
is predicted for the time between $t$ = 500.0 and $t$ = 999.9 in Fig.\ref{hindmarsh}(a) for the absence of noise and in Fig.\ref{hindmarsh}(b) for the presence of noise. The strength of noise is kept here as 0.01. The target is plotted by solid green line and the prediction is plotted by red dashed line.  The parameters were chosen as $a_{min}$ = 1, $a_{max}$= 2, $u_{min}$ = -1.2, $u_{max}$ = 1.8, $P$ = 5 and $m$ = 100. Also, the time step and the initial conditions for the training and prediction are taken as 0.1 and $(x^*,y^*,z^*)$ = (-1.6,4.0,2.75) and 0.1, respectively.  The training was done with 1000 data points between the times $t$ = 400.0 and $t$ = 499.9. The prediction is satisfactory as shown in the Figs.\ref{hindmarsh}(a) and (b), and the root mean square error for the predictions between $t$ = 500 and $t$ = 10000 (95000 data) is determined as 1.032240$\times$10$^{-3}$ for the prediction without noise and 1.464032$\times$10$^{-2}$ with noise.
	
\begin{table}
\begin{center}
\begin{tabular}{ |c|c|c|c| } 
\hline
Dynamical & ($a_{min},a_{max}$) & RMSE  & NRMSE \\ 
System	&  	 	&  ($\delta$=0) &( $\delta$=0) \\[0.5 ex]
\hline
&	(1, 2) & 1.541608$\times$10$^{-3}$ & 2.33887324$\times$10$^{-5}$ \\ [0.5 ex]
Lorenz	&(3.0, 3.4) & 8.99947$\times$10$^{-4}$  & 1.36539263$\times$10$^{-5}$  \\ [0.5 ex]
&(3.6, 4) & 7.703143$\times$10$^{-4}$ & 1.16872192$\times$10$^{-5}$ \\[0.5 ex]
\hline
&	(1, 2) & 4.660823$\times$10$^{-2}$ & 2.003829$\times$10$^{-3}$ \\ [0.5 ex]
R\"ossler	&(3.0, 3.4) & 4.772730$\times$10$^{-2}$     &   2.05197354$\times$10$^{-3}$      \\ [0.5 ex]
&(3.6, 4) &  4.744061$\times$10$^{-2}$      &   2.03963365$\times$10$^{-3}$       \\[0.5 ex]
\hline
Hindmarsh &	(1, 2) & 1.032240$\times$10$^{-3}$ & 3.77652143$\times$10$^{-4}$ \\ [0.5 ex]
-Rose &(3.0, 3.4) &  1.538142$\times$10$^{-3}$     & 5.62739888  $\times$10$^{-4}$\\ [0.5 ex]
&(3.6, 4) 	&   2.299932$\times$10$^{-3}$   	& 8.41439563$\times$10$^{-4}$ \\[0.5 ex]
\hline
Dynamical & ($a_{min},a_{max}$) & RMSE  & NRMSE \\ 
System	&  	 	&  ($\delta$=0.01) &( $\delta$=0.01) \\[0.5 ex]
\hline
&	(1, 2) & 7.168561$\times$10$^{-2}$ & 1.087678029$\times$10$^{-3}$ \\ [0.5 ex]
Lorenz	&(3.0, 3.4) & 7.286838$\times$10$^{-2}$  & 1.1057753916$\times$10$^{-3}$  \\ [0.5 ex]
&(3.6, 4) & 7.221521$\times$10$^{-2}$ & 1.0959847161$\times$10$^{-3}$ \\[0.5 ex]
\hline
&	(1, 2) & 5.376042$\times$10$^{-2}$ & 2.3116406$\times$10$^{-3}$ \\ [0.5 ex]
R\"ossler	&(3.0, 3.4) & 5.529051$\times$10$^{-2}$   &   2.37715542$\times$10$^{-3}$      \\ [0.5 ex]
&(3.6, 4) &  5.481263$\times$10$^{-2}$       &   2.3564095$\times$10$^{-3}$       \\[0.5 ex]
\hline
Hindmarsh &	(1, 2) & 1.464032$\times$10$^{-2}$ & 5.37781048$\times$10$^{-3}$ \\ [0.5 ex]
-Rose &(3.0, 3.4) &  1.378490$\times$10$^{-2}$    & 5.06357896$\times$10$^{-3}$  \\ [0.5 ex]
&(3.6, 4) 	&   1.440286$\times$10$^{-2}$   	& 5.26906593$\times$10$^{-3}$  \\[0.5 ex]
\hline
\end{tabular}
\caption{RMSE and NRMSE values for the three dynamical systems for different ranges of ($a_{min},a_{max}$) in the presence ($\delta$ = 0.01) and absence of noise. Here the values of $P$, $m$, $u_{min}$ and $u_{max}$ for the Lorenz, R\"ossler and Hindmarsh-Rose systems are kept as same as used in Figs.\ref{lorenz}, \ref{rossler} and \ref{hindmarsh}, respectively.}
\label{table:1}
\end{center}
\end{table}
	
In the nontemporal task the predictions were made with good accuracy for the parameter window $\triangle a$ = 0.1 since the error in the prediction becomes large when $\triangle a >$ 0.1. However, in temporal cases, where the nonlinear systems were predicted with $a_{max}$ = 2, $a_{min}$ = 1 and $\triangle a$ = 1, the parameter window can be increased up to 4. This implies that the temporal cases can be predicted accurately with the parameter $a$ in the periodic as well as chaotic ranges of the logistic map.  We have predicted the three nonlinear systems for the different ranges of parameter window (i) ($a_{min}$, $a_{max}$) = (1, 2), (ii) ($a_{min}$, $a_{max}$) = (3, 3.4) and (iii)($a_{min}$, $a_{max}$) = (3.6, 4) corresponding to the regions (logistic map) of period-1, period-2 cycle and chaos in the presence and absence of noise and the results are listed in Table.1. The RMSE values listed in Table.1 confirms the accuracy in the prediction of the three nonlinear systems for the larger parameter windows of $a$ i.e. in the periodic and chaotic regions of the logistic map.
	
\subsection{Simultaneous prediction of $y(t)$ and $z(t)$ from $x(t)$}
\begin{figure}
\centering
\includegraphics[width=0.9\linewidth]{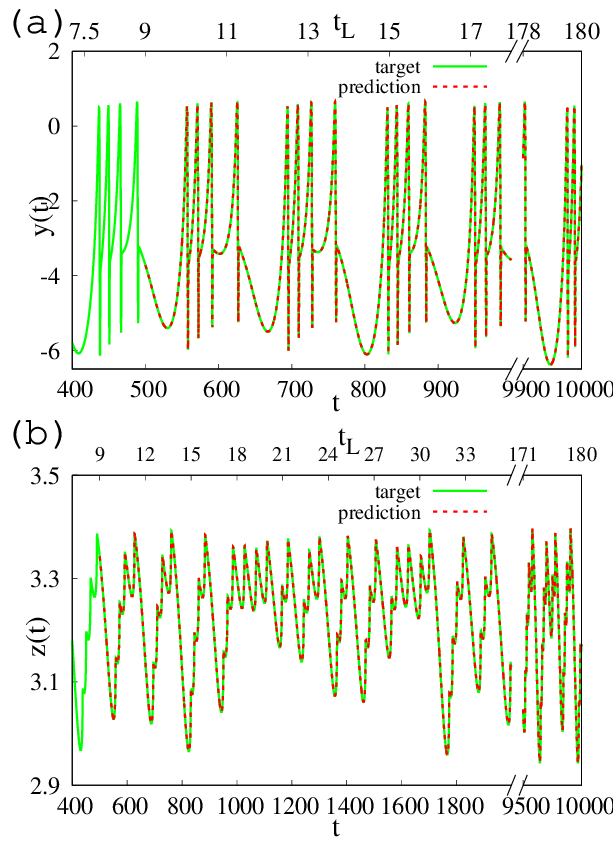}
\caption{Prediction of Hindmarsh-Rose system for $y(t)$ and $z(t)$ from the input $x(t)$ without noise. Here, $a_{min}$ = 3.6, $a_{max}$= 4, $u_{min}$ = -1.2, $u_{max}$ = 1.8, $P$ = 5 and $m$ = 100. }
\label{hindmarsh_yz}
\end{figure}

In the previous sections, only the $y(t)$ of the three nonlinear dynamical systems was predicted from the input $x(t)$.  Here, we intend to predict the variables $y(t)$ and $z(t)$ simultaneously from $x(t)$.  The procedure for this simultaneous prediction is given below: To obtain the weight matrix for this prediction the row vector $v$ in Eq.\eqref{tempW} can be replaced by a matrix $[v^y,v^z]^T$ with two row vectors $v^y$ and $v^z$ having output elements $[v_1^y,v_2^y,.....,v_L^y]$ and $[v_1^z,v_2^z,.....,v_L^z]$ corresponding to $y(t)$ and $z(t)$, respectively. The reservoir state vector matrix $R$ is constructed from the input vector 	$u=[u_1,u_2,.....,u_L]$ corresponding to $x(t)$ as before. After finding the  weight matrix $W_t$ (with size $2\times mP$), both the variables $y(t)$ and $z(t)$ are predicted for the input $u'=[u_2,u_3,.....,u_{L+1}]$ by the relation given by Eq.\eqref{tempV} for the output $V ~=~ [V^y,V^z]^T$. Here $V^y = [V^y_2,V^y_3,.....,V^y_{L+1}]$ and $V^z = [V^z_2,V^z_3,.....,V^z_{L+1}]$ are the output vectors corresponding to $y(t)$ and $z(t)$, respectively. The $y(t)$ and $z(t)$ variables of the Hindmarsh-Rose system are predicted  from the input $x(t)$ in the absence of noise and plotted in Fig.\ref{hindmarsh_yz} for $a_{min}$ = 3.6, $a_{max}$= 4, $u_{min}$ = -1.2, $u_{max}$ = 1.8, $P$ = 5 and $m$ = 100. The figure confirms that the prediction is good and the RMSEs are determined to be 2.922806$\times$10$^{-4}$ and 3.059959$\times$10$^{-4}$ for $y(t)$ and $z(t)$, respectively.	
We have also predicted the self prediction of $x(t)$ of the R\"ossler system from the same input $x(t)$ by logistic map and trigonometric series function(for details see Appendix).

\subsection{Self prediction (or) Closed-loop prediction}
\begin{figure}[!h]
\centering
\includegraphics[width=0.9\linewidth]{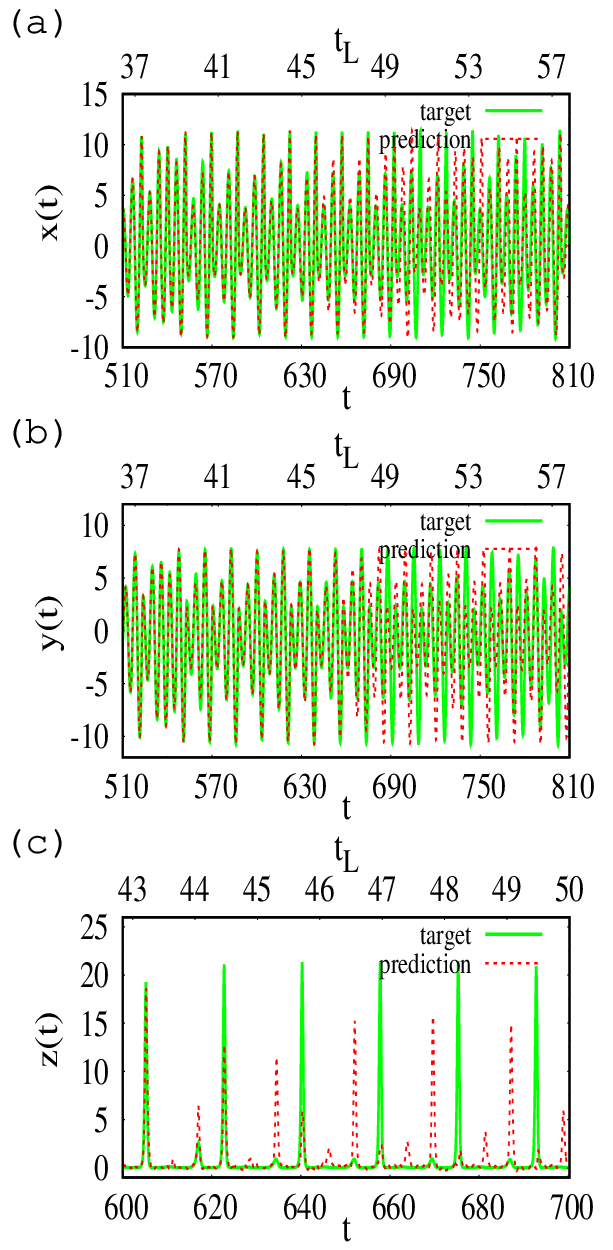}
\caption{Prediction of R\"ossler system for (a) $x(t)$ when $P$ = 20, $a_{min}$ = 3, $a_{max}$ = 3.5, $u_{min}$ = -10, $u_{max}$ = 15, $m$ = 100 and $L$ = 5000, (b) $y(t)$ when $P$ = 35, $a_{min}$ = 3.6, $a_{max}$ = 4, $u_{min}$ = -15, $u_{max}$ = 15, $m$ = 100 and $L$ = 5000, and (c) z(t) when $P$ =2, $a_{min}$ = 1, $a_{max}$ = 2, $u_{min}$ = -1, $u_{max}$ = 25, $m$ = 1000 and $L$ = 5000, from the inputs $x(t)$, $y(t)$ and $z(t)$, respectively, for the strength of noise 0.01.}
\label{selfpred}
\end{figure}
Next, by making use of the above procedure, the time-series of $x(t)$, $y(t)$ and $z(t)$ of the R\"ossler system are predicted by using the same $x(t)$, $y(t)$ and $z(t)$ variables, respectively, as input and the results are presented in Figs.\ref{selfpred}(a), (b) and (c), respectively, with the strength of noise $\delta$ = 0.01. Here the target is plotted as green line and the prediction is plotted as red dashed line. For $x(t)$ and $y(t)$, the training is made with 5000 number of data points collected between $t$ = 10.0 and $t$ = 509.9 with the time-space 0.1 and the prediction is made from $t$ = 510.1 to $t$ = 810.0 with the weight matrix $W_t$ obtained at $t$ = 510.0. The weight matrix $W_t$ is determined by $ W_t = v R^{-1} $ using the reservoir state vector matrix $R$ formed with the inputs from $t$ = 10.0 to $t$ = 509.9 and the outputs $v$ from $t$ = 10.1 to $t$ = 510.0. In Fig.\ref{selfpred}(a), a close prediction with the target up to $t$ = 690 is observed for $x(t)$  for $P$ = 20, $a_{min}$ = 3, $a_{max}$ = 3.5, $u_{min}$ = -10 and $u_{max}$ = 15.  The RMSE and NRMSE are determined as 4.272064$\times$10$^{-1}$ and 1.701691$\times$10$^{-2}$, respectively, for the prediction between $t$ = 510.1 and $t$ = 610. Similarly, as may be observed from Fig.\ref{selfpred}(b), the variable $y(t)$ is also closely predicted up to $t$ = 680 with $P$ = 35, $a_{min}$ = 3.6 and $a_{max}$ = 4, $u_{min}$ = -15 and $u_{max}$ = 15. The RMSE and NRMSE are measured as 4.350448$\times$10$^{-1}$ and 1.946701$\times$10$^{-2}$, respectively, for the prediction between $t$ = 510.1 and $t$ = 610. The $z(t)$ variable is predicted from $t$ = 600.1 to $t$ = 700 with $P$ = 2, $a_{min}$ = 1 and $a_{max}$ = 2, $u_{min}$ = -1 and $u_{max}$ = 25 from the training of 5000 data points between $t$ = 100.0 and $t$ = 599.9  and  the result is plotted in Fig.\ref{selfpred}(c), where we can see that the prediction matches well with the target up to $t$ = 630 and then deviation starts to appear. For this prediction, the memory stack in the column vector of the reservoir state vector matrix $R$ is formed with 1000 entries (i.e. $m$ = 1000) with the memory weights 0.001, 0.002,...,1.0, instead of $m$ = 100 considered for predicting $x(t)$ and $y(t)$. The RMSE and NRMSE are determined as 1.147877 and 1.520118$\times$10$^{-1}$ for the prediction between $t$ = 600.1 and $t$ = 630.0,.
\subsection{Higher-dimensional system}
\begin{figure}[!h]
	\centering
	\includegraphics[width=1\linewidth]{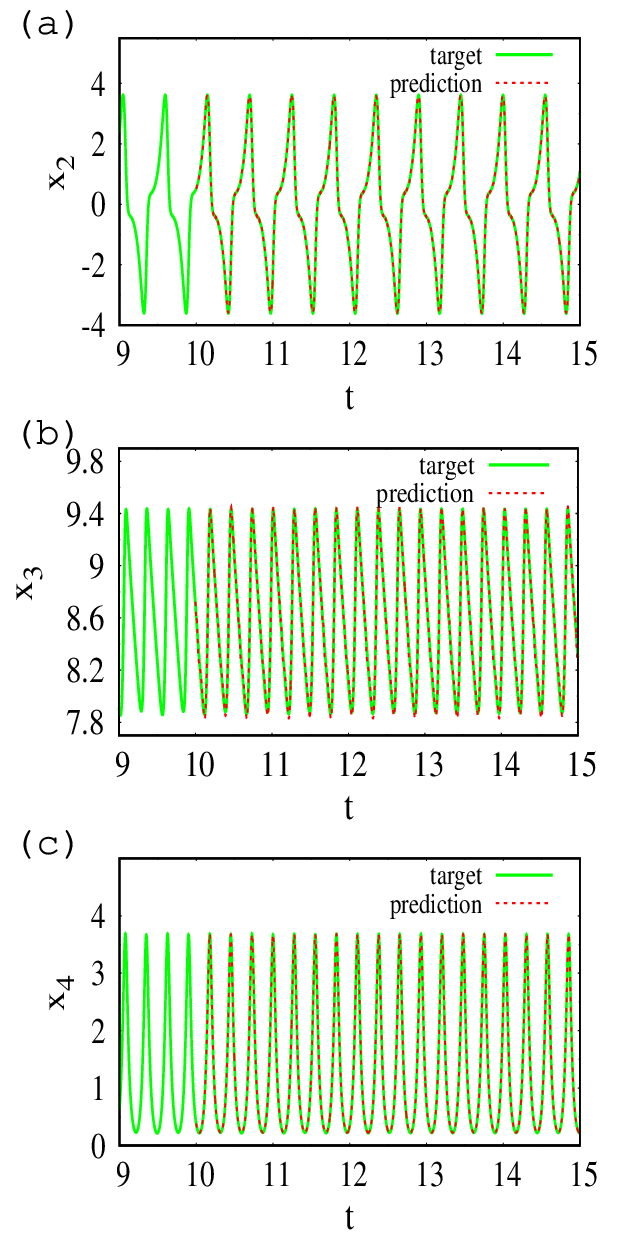}
	\caption{Prediction of the four dimensional system for the variable (a) $x_2$ from the input $x_1$, (b) $x_3$ from the input $x_2$ and (c) $x_4$ from the input $x_3$. Here $P$ = 3, $L$ = 1000, $a_{min}$ = 3.6, $a_{max}$ = 3.9, $u_{min}$ = -6, $u_{max}$ = 10 and $m$ = 100.}
	\label{higherdim}
\end{figure}
The applicability of our methodology of reservoir is further verified for the four dimensional chaotic system considered by Qi $et~al$ \cite{qi} as given below:
\begin{align}
	&\dot{x}_1 = a(x_2-x_1)+x_2 x_3 x_4,\nonumber\\
	&\dot{x}_2 = b(x_1+x_2)-x_1 x_3 x_4,\nonumber\\
	&\dot{x}_3 = -c x_3 + x_1 x_2 x_4,\nonumber\\
	&\dot{x}_4 = -d x_4 +x_1 x_2 x_3\label{fourdim}.
\end{align}
In Figs.\ref{higherdim}(a), (b) and (c) the variables $x_2$, $x_3$ and $x_4$ are predicted from the inputs $x_1$, $x_2$ and $x_3$, respectively for the duration of time from $t$ = 10 to $t$ = 15.0. For the given inputs the above system is numerically solved with the initial conditions (0.1,0.2,0.3,0.4) for $(x_1,x_2,x_3,x_4)$ and for $a$ = 35, $b$ = 10, $c$ = 1 and $d$ = 25. The step size for time is kept as 0.001. The training is done from $t$ = 9.000 to 9.999 with $L$ = 1000 data points and the prediction is shown from 10.0 to 15.0. The values of $P$, $a_{min}$, $a_{max}$, $u_{min}$ and $u_{max}$ are kept as 3, 3.6, 3.9, -6 and 10, respectively. Here, the strength of noise is taken as zero. From Figs.\ref{higherdim} we can observe that the predictions on $x_2$, $x_3$ and $x_4$ are exact. The accuracy is maintained for long time and the $(RMSE,NRMSE)$ are determined for the time period between $t$ = 10.0 to $t$ = 100.0 as (3.766656$\times 10^{-2}$,1.074041$\times 10^{-2}$), (2.549922$\times 10^{-2}$,0.09876749$\times 10^{-2}$) and (2.098983$\times 10^{-3}$,1.65217$\times 10^{-3}$), respectively.

\section{Performance of RC}
\begin{figure}[!h]
	\centering
	\includegraphics[width=0.9\linewidth]{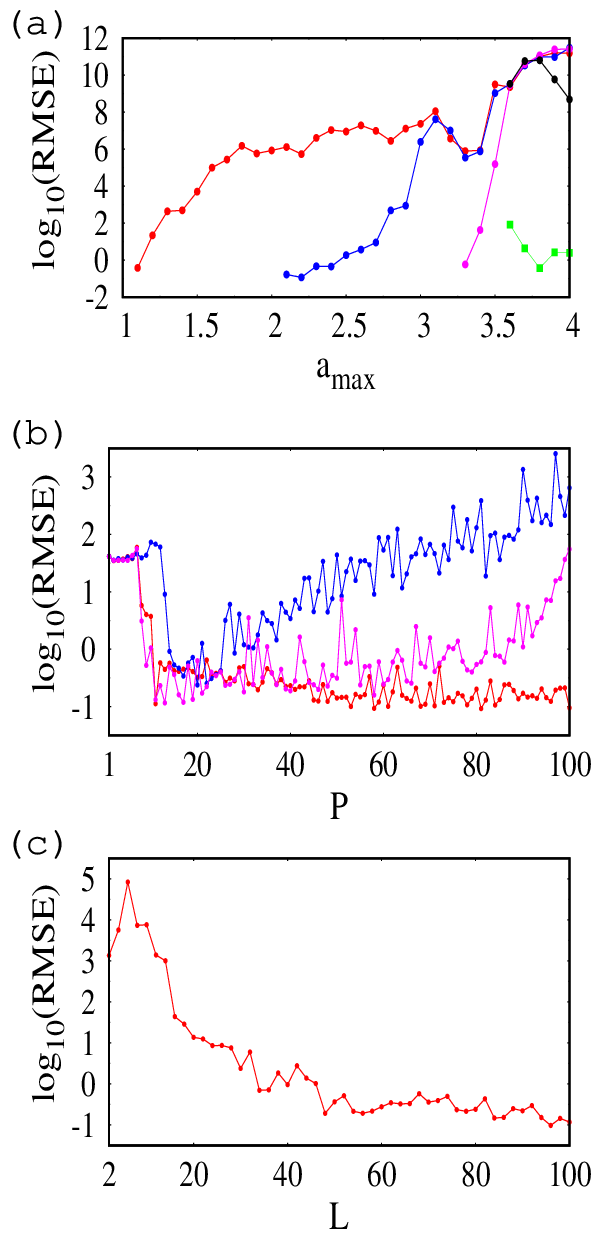}
	\caption{log$_{10}$(RMSE) of the polynomial prediction task against (a) $a_{max}$ when $P$ = 100 and $L$ = 100, (b) logistic map iteration length $P$ when $a_{min}$ = 2.1, $a_{max}$=2.2 (red), $a_{min}$ = 3.57, $a_{max}$ = 3.58 (blue) and $a_{min}$ = 3.8, $a_{max}$ = 3.9 (magenta) and (c) input length $L$ when $P$=100 and $a_{min}$ = 2.1, $a_{max}$=2.2. Here, $u_{min}$ = -3.2, $u_{max}$ = 3.2 and the noise strength is 0.1.}
	\label{performpoly}
\end{figure}

The performance of the reservoir against the hyper-parameters $a_{min}$, $a_{max}$, $P$ and $L$ are discussed here for both the temporal and nontemporal tasks. In Figs.\ref{performpoly}(a-c), for the prediction of the seventh degree polynomial function discussed in Sec.II, the logarithmic values of RMSE are plotted against the upper limit of bifurcation value of the logistic map $a_{max}$, the number of iterations of the logistic map $P$ and the length of the inputs $L$, respectively for the noise strength 0.1, $L$ = 100 and $u_{min}$ = -3.2, $u_{max}$ = 3.2.  In Fig.\ref{performpoly}(a) $log_{10}(RMSE)$ is plotted for different values of $a_{min}$, 1.0(red), 2.0(blue), 3.2(magenta) and 3.5(black) for $P$ = 100 and 3.5(green) for $P$ = 20. From Fig.\ref{performpoly}(a), we can observe that if the value of `$a$' is in the nonchaotic region, the error in the prediction increases when the difference between $a_{max}$ and $a_{min}$ increases. If `$a$' is in the chaotic region, the error becomes large when $P$ is large and can be reduced by decreasing the value of $P$.  In Fig.\ref{performpoly}(b), the $log_{10}(RMSE)$ is plotted against the number of iterations $P$ for ($a_{min},a_{max}$): (2.1,2.2), (3.57,3.58) and (3.8,3.9), by the  red, blue and magenta colour lined points, respectively, while $L$ = 100. From Fig.\ref{performpoly}(b) we can identify that if `$a$' is in the nonchaotic region, i.e. $a_{min}$=2.1, $a_{max}$=2.2, the error reduces drastically once the value of $P$ increases above 10, whereas in the region which is identified as edge of chaos($a_{min}$=3.57, $a_{max}$=3.58) or in the chaos region ($a_{min}$=3.8, $a_{max}$=3.9), the error reduces initially with $P$ and then increases considerably while increasing the value of $P$. In Fig.\ref{performpoly}(c), where the $log_{10}(RMSE)$ is plotted against the number of input length $L$ for $P$ = 100, $a_{min}$ = 2.1 and $a_{max}$ = 2.2, we observe that the error can be minimized when the length of the input $L$ is increased above 40.

\begin{figure}[!h]
	\centering
	\includegraphics[width=0.9\linewidth]{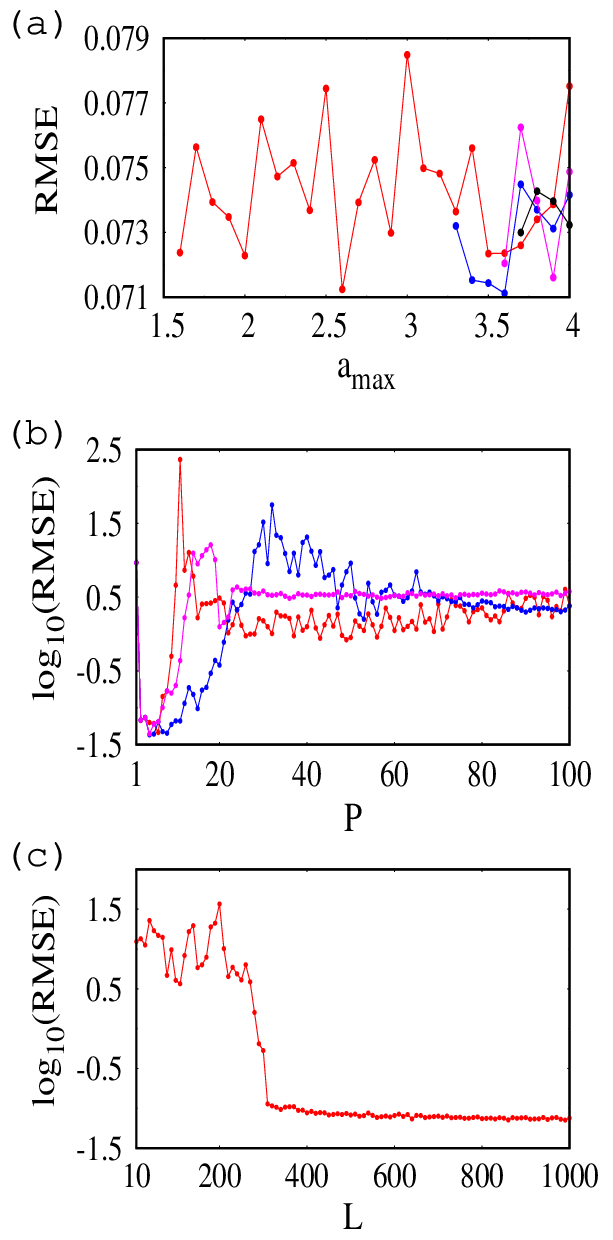}
	\caption{Prediction of $y$ from $x$ from the Lorenz system for (a) RMSE against $a_{max}$ when $P$ = 3 and $L$ = 1000, (b) log$_{10}$(RMSE) against $P$ for $L$ = 1000 when (i) $a_{min}$ = 1, $a_{max}$ = 2 (red), (ii) $a_{min}$ = 3.57, $a_{max}$ = 3.58 (blue) and (iii) $a_{min}$ = 3.8, $a_{max}$ = 3.9 (magenta), and (c) log$_{10}$(RMSE) against $L$ when $P$ = 3 and $a_{min}$ = 1, $a_{max}$ = 2. Here, $u_{min}$ = -17, $u_{max}$ = 17 and the noise strength is 0.01.}
	\label{performlorenz}
\end{figure}

To investigate the performance of the reservoir for the temporal task, Fig.\ref{performlorenz}(a) is plotted for the RMSE (for the prediction of $y$ from $x$ of the Lorenz system) against $a_{max}$ for different values of $a_{min}$ = 1.5(red), 3.2(blue), 3.5(magenta) and 3.6(black) while $P$ = 3, $L$ = 1000,  $u_{min}$ = -17 and $u_{max}$ =  17. From Fig.\ref{performlorenz}(a) we can understand that the error is small and there is no impact on RMSE by changing the values of $a_{min}$ and $a_{max}$. Figs.\ref{performlorenz}(b) and (c) are plotted for $log_{10}(RMSE)$ against $P$ when $L$ = 1000 and $L$ when $P$=3, respectively for  $u_{min}$ = -17 and $u_{max}$ =  17. In Fig.\ref{performlorenz}(b) the red, blue and magenta colour lined points correspond to the nonchaotic region ($a_{min}$ = 1, $a_{max}$ = 2), edge of chaos ($a_{min}$ = 3.57, $a_{max}$ = 3.58) and chaotic region ($a_{min}$ = 3.8, $a_{max}$ = 3.9), respectively. From Fig.\ref{performlorenz}(b) we can observe that irrespective to the region of $a$ the error in the prediction becomes very small at low values of $P$ and gets enhanced slightly for large values of $P$. Fig.\ref{performlorenz}(c), which has been plotted for $a_{min}$ = 1 and $a_{max}$ = 2, confirms that the error drastically reduces when the length of the input $L$ increases above 300. 

\section{Conclusion}
We have successfully demonstrated that both temporal and nontemporal tasks can be predicted well by constructing a reservoir using the well-known nonlinear map, that is the logistic map, or the map with a simple trigonometric series (when needed) in reservoir computing. The logistic map is used to transform the input into a higher dimensional system  by constructing virtual nodes. We have predicted a seventh order polynomial for nontemporal tasks and three nonlinear systems, namely the Lorenz, R\"ossler, and Hindmarsh-Rose oscillators, for temporal tasks with logistic map. The polynomial is accurately predicted in the absence and presence of noise, with the root mean square error 3.475412$\times$10$^{-3}$ and 6.152246$\times$10$^{-2}$, respectively. Also, we have predicted the three nonlinear dynamical systems for the variable $y(t)$ from the input $x(t)$ in the presence and absence of noise with a strength of 10$^{-2}$. The root mean square errors for the Lorenz, R\"ossler, and Hindmarsh-Rose systems, are determined as 1.541608$\times$10$^{-3}$, 4.660823$\times$10$^{-2}$, and 1.03224$\times$10$^{-3}$, respectively, in the absence of noise. The same errors are determined as 7.168561$\times$10$^{-2}$, 5.376042$\times$10$^{-2}$, and 1.464032$\times$10$^{-2}$, respectively, in the presence of noise strength 10$^{-2}$. The low values of the root mean square error for both the temporal and nontemporal tasks confirm that the accuracy of the above method is quite satisfactory and that it can be employed as an effective reservoir in reservoir computing. We have also, predicted the time-series of $x$, $y$ and $z$ variables of the R\"ossler system under the closed-loop task i.e. without taking the true values of other inputs as input, and have obtained the RMSE as 4.272064$\times$10$^{-1}$, 4.350448$\times$10$^{-1}$ and 1.147877, respectively. The method has also been demonstrated well for the four-dimensional system (Eq.\eqref{fourdim}). The parameter window for the input is increased by constructing a simple trigonometric series for time multiplexing the input with virtual nodes. The time series of $x(t)$ of the R\"ossler system is predicted with good accuracy by the logistic map as well as with the trigonometric series.  Also, the self prediction of the remaining two state variables  $y(t)$ and $z(t)$ of the same system and  the three state variables of the other two nonlinear systems are verified (results not shown here) by the logistic map and trigonometric series.

	
\section*{Appendix}
\subsection*{A. Multiplexing with a Simple Trigonometric Series}
In Sec. II the polynomial was predicted with the parameter window $\triangle a$ = 0.1 and it was mentioned that the error will become large when the range is increased further. If there is a requirement for supplying large number of data as input the parameter window should be large enough to accommodate the inputs and it will be preferable if the value of $\triangle a$ is  $\geq$0.5. 
\begin{figure}[h]
\centering
\includegraphics[width=1\linewidth]{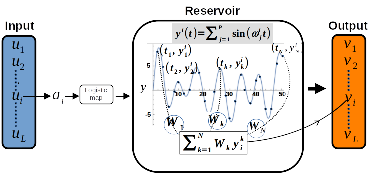}
\caption{The schematic diagram for training with trigonometric function. The input $u_i$ is linearly transformed into $a_i$ and then is supplied into logistic map to form a time varying function. $V_i$, $i=1,2,..,L$ is the output generated from the reservoir. The $W_k,~k=1,2,...,P$ are the components of the weight matrix $W$.  The black colour dotted lines inside the reservoir indicate the internal computations of the reservoir.} 
\label{sinres}
\end{figure}
Here, we discuss a methodology to increase the size of the parameter window to predict the polynomial in the nontemporal task by constructing a simple trigonometric function formed by the iterated values of logistic map  $\omega_1^i,\omega_2^i,...,\omega_P^i$ corresponding to each input $u_i$.  The input $u_i$ is then multiplexed in time via $a_i$ by a simple trigonometric series function given by (see Fig.\ref{sinres})
	\begin{align}
		y^{i}(t) = \sum_{k=1}^{k=P} \sin(\omega^i_k t). \tag{A1} \label{y}
	\end{align}
	From Eq.\eqref{y}, one can collect $N$ data points $y_1^{i},y_2^{i},....,y_N^i$ for $y^i(t)$ at $t$ = $t_j$, where $j$ = 1, 2,...,$N$ corresponding to $u_i$. Here the step size ($t_{j+1}-t_j)$ is given by $t_N/N$, and $t_0=0$ and $t_N$ are the initial and final values of $t$, respectively.  $N$ is the total number of data points we intend to collect corresponding to each $u_i$ or $a_i$ for the prediction. Here, $t_N$ and $N$ are the hyper-parameters.  $t_0$ is taken as 0 for this entire work. A reservoir state vector(column vector) ${Y}_{i}$ corresponding to $a_i$ and $u_i$ is formed with size ($N\times 1$) from the data points $y_1^{i},y_2^{i},....,y_N^{i}$ as $	{Y}_{i} = [ y_{i}^1,y_{i}^2,....,y_{i}^N]^T$. After constructing the reservoir state vectors $Y_{1},Y_{2},....,Y_{L}$ corresponding to $u_1,u_2,...,u_L$ ($L$ inputs) using Eq.\eqref{y}, they are stacked together to from a reservoir state vector matrix $R = [Y_{1},Y_{2},....,Y_{L}]$ with the size $N\times L$ as
	\begin{align}
		R = 
		\begin{bmatrix}
			y_{1}^1 & y_{1}^2 & y_{1}^3 & ......&y_{1}^L\\~\\
			y_{2}^1 & y_{2}^2 & y_{2}^3 & ......&y_{2}^L\\~\\
			y_{3}^1 & y_{3}^2 & y_{3}^3 & ......&y_{3}^L\\
			...&...&...&...&...\\
			...&...&...&...&...\\
			y_{N}^1 & y_{N}^2 & y_{N}^3 & ......&y_{N}^L
		\end{bmatrix}. \tag{A2}
	\end{align}
	From the reservoir state vector matrix $R$ and the output vector $v$ the weight matrix $W_{nt}$ for the nontemporal task is obtained by using Eq.\eqref{W} and the output is predicted as given by Eq.\eqref{vWR}.  The polynomial $f(x)$ is predicted based on this procedure and plotted in Fig.\ref{fig5} without noise for the parameter window $\triangle a$ = 0.5. The parameters are $a_{min}=3.0 $ and $ a_{max}=3.5$, $P$ = 5, $N$ = 100, $t_0$ = 0 and $t_N$ = 10. The solid blue line is for target and the red open circle is for prediction. The RMSE is determined as 6.896466$\times$10$^{-2}$. 
	\begin{figure}
		\centering
		\includegraphics[width=1\linewidth]{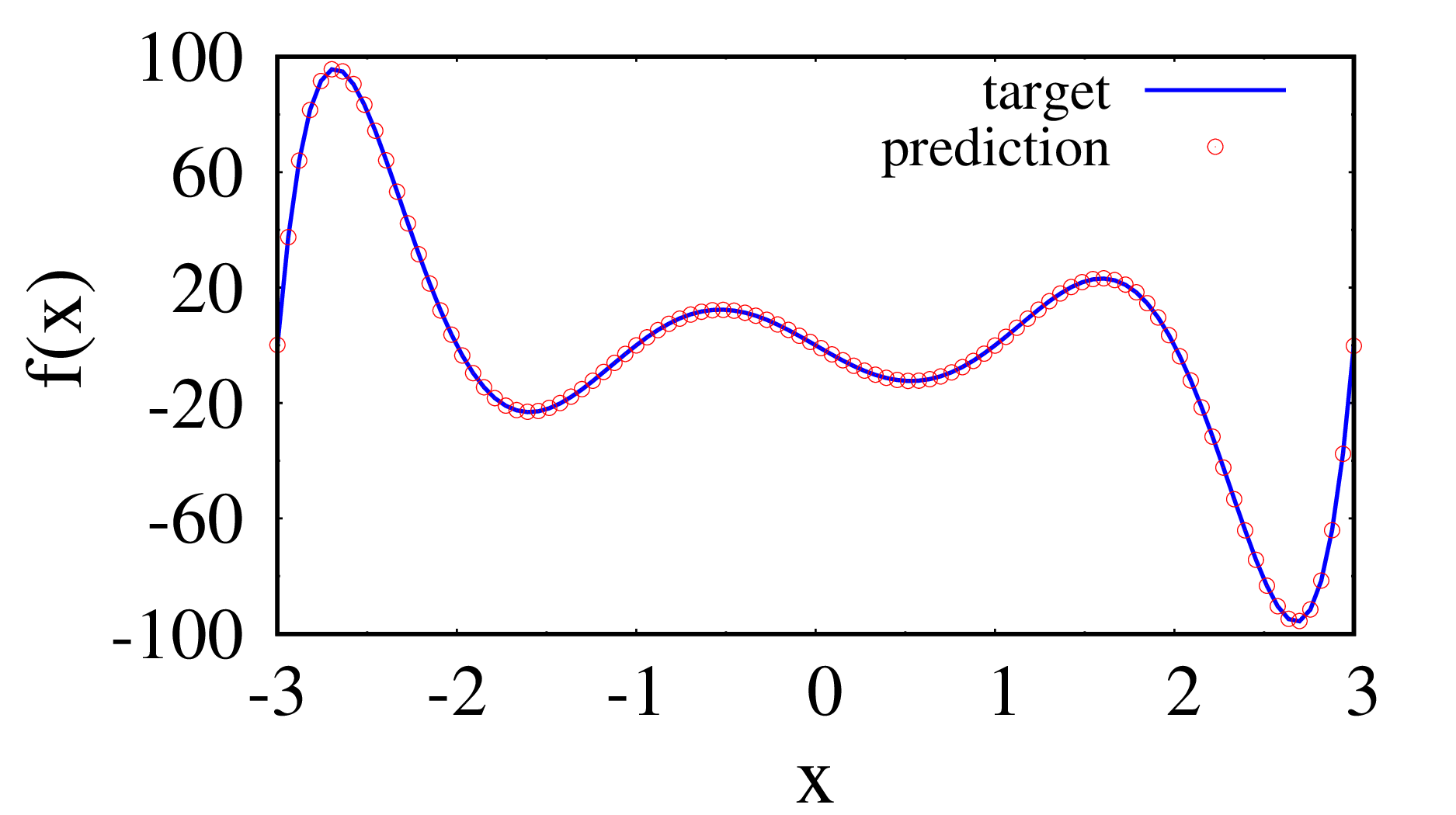}
		\caption{The prediction of polynomial $f(x)$ = $(x-3)(x-2)(x-1)x(x+1)(x+2)(x+3)$ between -3 and +3 in the absence of noise with $a_{min}$ = 3 and $a_{max}$ = 3.5. Here $L$ = 100, $P$ = 100,  $u_{min}$ = -3 and $u_{max}$ = 3.}
		\label{fig5}
	\end{figure}
	\begin{figure}[!h]
	\centering
	\includegraphics[width=0.8\linewidth]{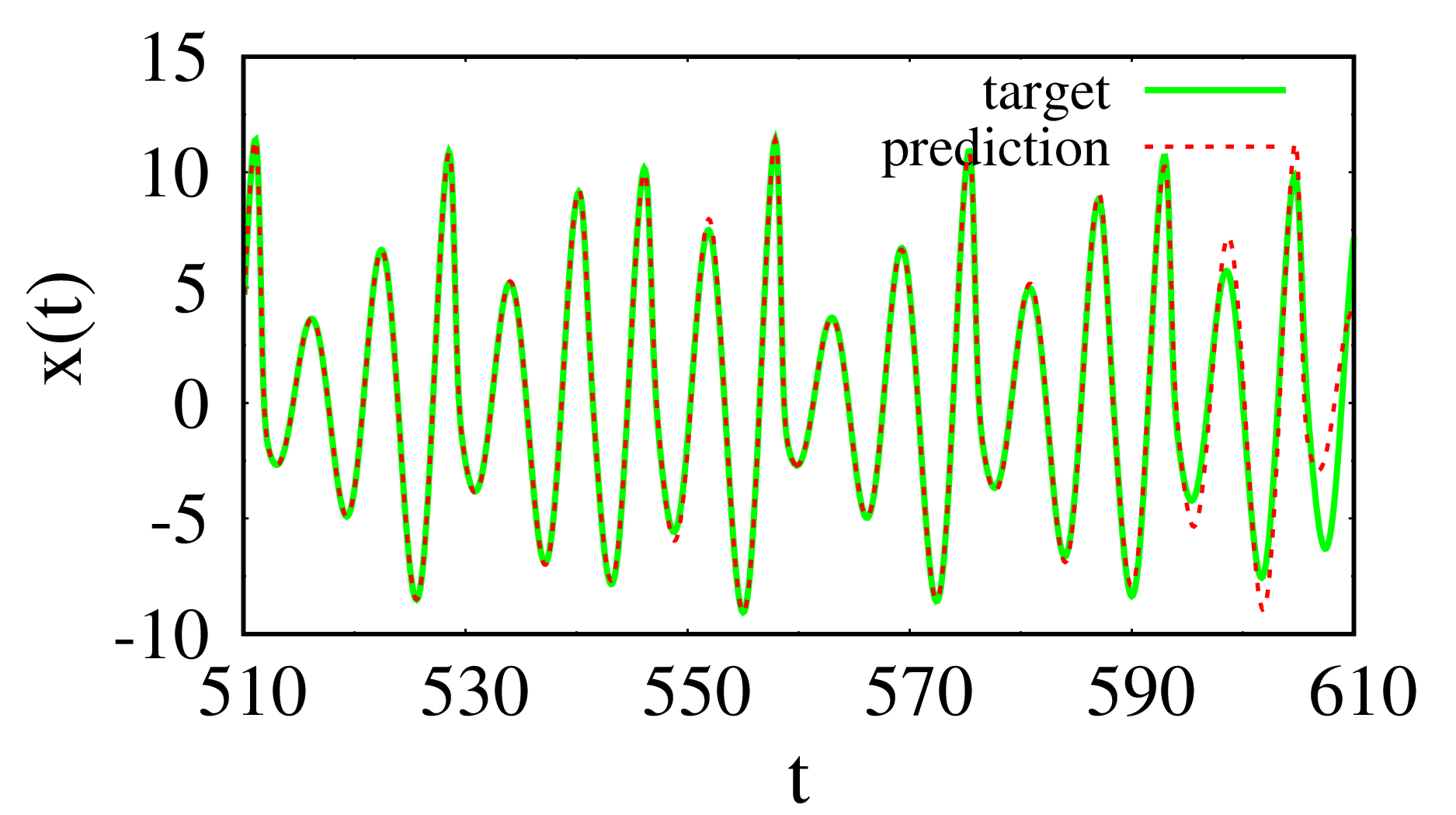}
	\caption{Prediction of R\"ossler system for $x(t)$ from the input $x(t)$ for the strength of noise 0.01 by (a) trigonometric function with $P$ = 5, $N$ = 100, $t_N$ = 10, $a_{min}$ = 1 and $a_{max}$ = 3.5.}
	\label{trigself}
\end{figure}
This procedure of trigonometric series can also be used to predict temporal tasks with reservoir matrix constructed by $R = [Z_1,Z_2,...,Z_L]$ similar to the construction procedure used for temporal task discussed in Sec.III above. Here the reservoir state vector $Z_{i}$ corresponding to the input $u_i$ is given by $Z_i = [g_0 Y_{{i-m}},.........,g_{m-1} Y_{{i-1}},g_m Y_{i}]^T$ along with $m$ previous inputs, where $Y_i = [y_1^i,y_2^i,...,y_N^i]^T$. The data points $y_1^i,y_2^i,...,y_N^i$ for the input $u_i$ are obtained from Eq.\eqref{y} instead of $\omega_1^i,\omega_2^i,...,\omega_P^i$ obtained from logistic map. 
\begin{figure}[!h]
	\centering
	\includegraphics[width=1\linewidth]{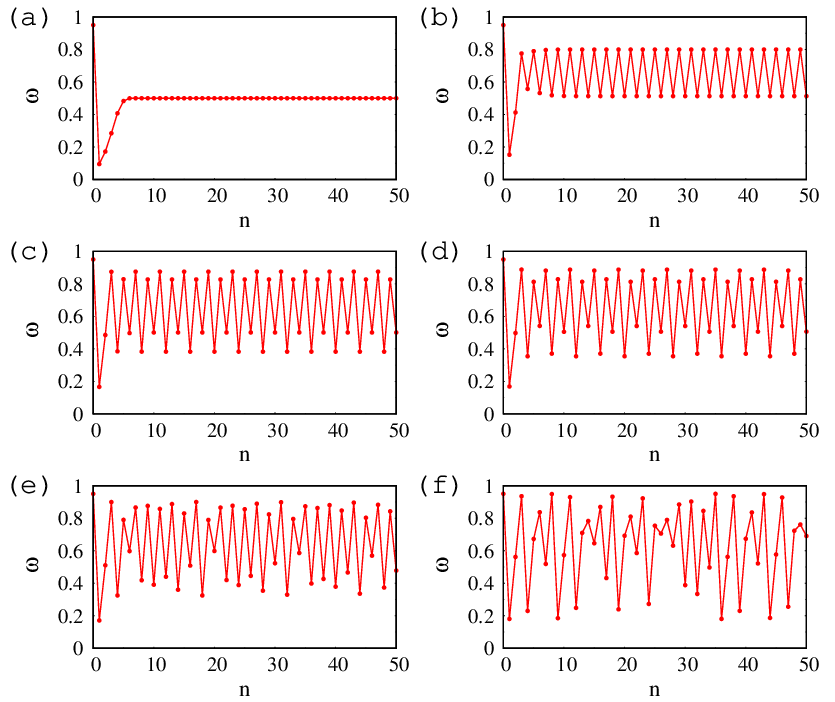}
	\caption{The iterated values of $\omega_n$ from logistic map equation Eq.\eqref{logistic} when (a) $a$ = 2.0, (b) $a$ = 3.2, (c) $a$ = 3.5, (d) $a$ = 3.55, (e) $a$ = 3.6 and (f) $a$ = 3.8 while $\omega_0$ = 0.95.}
	\label{fig15}
\end{figure}

By making use of this procedure, the time series of $x(t)$ of the R\"ossler system is predicted by using the same $x(t)$ as an input. The training is made with the 5000 number of data points collected between $t$ = 10.0 and $t$ = 509.9 with the time-space 0.1 and the prediction is made from $t$ = 510.0 to $t$ = 609.9 with the weight matrix $W_t$ obtained at $t$ = 509.9. The weight matrix $W_t$ is determined by $ W_t = v R^{-1} $ using the reservoir state vector matrix $R$ formed with the inputs from $t$ = 10.0 to $t$ = 509.9 and the outputs $v$ from $t$ = 10.1 to $t$ = 510.0. The close prediction with the target is shown in Fig.\ref{trigself} where the target is plotted by green line and the prediction is plotted by red dashed line for $N$ = 100, $t_N$ = 10, $P$ = 5, $a_{min}$ = 1, $a_{max}$ = 3.5 along with the noise strength 0.01.  The RMSE and NRMSE are determined as 0.735332 and 2.7943088$\times 10^{-2}$, respectively.

\subsection*{B. Dynamical Behaviour of the Logistic Map\cite{lak}}
The iterated values of logistic map equation $\omega_{n+1} = a \omega_n(1-\omega_n)$, where $n$ = 0, 1, 2, ...,P, are plotted in Fig.\ref{fig15} for different values of $a$ = 2.0, 3.2, 3.5, 3.55, 3.6 and 3.8 while $P$ = 50. As we can see from the Fig.\ref{fig15}(a), after transients the variable $\omega_n$ exhibits fixed point values while $a$ = 2.0. Figs.\ref{fig15}(b), (c) and (d) exhibit 2-periodic, 4-periodic, 8-periodic oscillations, respectively, corresponding to $a$ = 3.2, 3.5 and 3.55. Figs.\ref{fig15}(e) and (f) exhibit chaotic nature when $a$ = 3.6 and 3.8.

\section*{Acknowledgments}
M.L. wishes to thank the Department of Science and Technology for the award of a DST-SERB National Science Chair  under Grant No. NSC/2020/00029 in which R.A. and M.S. are supported by Research Associateships.\\~\\

\end{document}